\documentclass[11pt]{article}
\usepackage{graphicx}

\newcommand{\be}[1]{\begin{equation} #1 \end{equation}}
\newcommand{\abs}[1]{\left| #1 \right|}
\newcommand{\paren}[1]{\left( #1 \right)}
\newcommand{\onehalf}{\frac{1}{2}}
\newcommand{\dis}{\displaystyle}
\def\etal{{\it et al.\ }}

\begin{document}


{\hfill 28 December 2007}

\begin{center}
\textbf{\Large Ruppeiner Geometry of Black Hole Thermodynamics}\\
\vspace{1cm}
\normalsize {\bf Jan E. {\AA}man}\footnote{ja@physto.se},
{\bf Narit Pidokrajt}\footnote{narit@physto.se} \\[.2cm]
\emph{Department of Physics \\
Stockholm University\\
SE-106 91 Stockholm\\
Sweden} \\[.4cm]
\end{center}

\begin{abstract}

The Hessian of the entropy function can be thought of as a metric
tensor on state space. In the context of thermodynamical
fluctuation theory Ruppeiner has argued that the Riemannian geometry
of this metric gives insight into the underlying statistical
mechanical system; the claim is supported by numerous examples.  We
study these geometries for some families of black holes and find that
the Ruppeiner geometry is flat for Reissner--Nordstr\"om black holes
in any dimension, while curvature singularities occur for the Kerr
black holes. Kerr black holes have instead flat Weinhold curvature.

\end{abstract}

We are here not studying the spacetimes of black holes as such.
Instead we examine entropy $S(M,Q,J)$ or mass $M(S,Q,J)$,
where $Q$ is charge and $J$ angular momentum.  The Ruppeiner metric
\cite{ruppeiner2} is
the Hessian of negative entropy: $g^{R}_{ij} = -
\partial_i\partial_jS(M, N^a)$ while the Weinhold metric
\cite{weinhold} is the
Hessian of $M$: $g^{W}_{ij} = \partial_i\partial_jM(S,N^a)$.
The two metrics are conformally related $ds^2 = g^{R}_{ij}dM^idM^j =
\frac{1}{T}g^W_{ij}dS^idS^j$ where $M^i = (M, N^a)$, $S^i = (S, N^a)$
and $T = \frac{\partial M}{\partial S}$ is the temperature.

\section{The Reissner--Nordstr\"om black hole}

The event horizon is given by
\begin{equation}
||{\xi}||^2 = - \left(1 - \frac{2M}{r} + \frac{Q^2}{r^2}\right) = 0\ ,
\end{equation}
where $r$ is chosen so that the area of a sphere at constant
$r$ equals $4{\pi}r^2$. There are two roots $r_+$ and
$r_-$. We find that $M = \frac{1}{2}(r_+ + r_-)$, $Q^2 = r_+r_-\ $,
$r_+ =\! M + \sqrt{M^2 -Q^2}
=\! M\,\left(1 + \sqrt{1-\frac{Q^2}{M^2}}\right)$.
The entropy is
$S = \frac{k}{4}A = k{\pi}r_+^{\,\,2} = r_+^{\,\,2}\ $, and we set
$k = 1/{\pi}$. Solving for $M$ the fundamental
thermodynamical relation is $M = \frac{\sqrt{S}}{2}\left( 1 +
\frac{Q^2}{S}\right)$. The temperature is 
$T = \frac{\partial M}{\partial S} = \frac{1}{4\sqrt{S}}\left( 
1 - \frac{Q^2}{S}\right)$. In its natural coordinates the Weinhold
metric becomes \cite{ourpaper1}

\begin{equation}
ds^2_W = \frac{1}{8S^{\frac{3}{2}}}
\bigg( - \left(1 - \frac{3Q^2}{S}\right)dS^2 - 8QdQdS + 8SdQ^2\bigg) \ .
\end{equation}

\noindent Introducing new coordinate $u = \frac{Q}{\sqrt{S}}$ ; $- 1
\leq u \leq 1$ we now find that

\begin{equation} ds^2_W = \frac{1}{8S^{\frac{3}{2}}}\left( - (1-u^2)dS^2
+ 8S^2du^2\right) \ . \end{equation}

\noindent In these coordinates the Ruppeiner metric is given by 

\begin{equation}
ds^2_R = \frac{1}{T}ds^2_W = - \frac{dS^2}{2S} + \frac{4Sdu^2}{1-u^2} 
\ . \end{equation}

\noindent This metric is flat! To see this introduce new coordinates 
${\tau} = \sqrt{2S}$, $\sin{\frac{\chi}{\sqrt{2}}} = u$.
The Ruppeiner metric now takes Rindler form 
$ds^2_R = - d{\tau}^2 + {\tau}^2d{\chi}^2$,
$\tau$ and $\chi$ are Rindler coordinates on the forward
light cone in Min\-kow\-ski space. In Fig. 1 we use the
inertial coordinates $t = \tau \cosh{\chi}$, $x = \tau \sinh{\chi}$.

\begin{figure}[h]
\begin{center}
\begin{tabular}{cc}

\includegraphics[width=45mm]{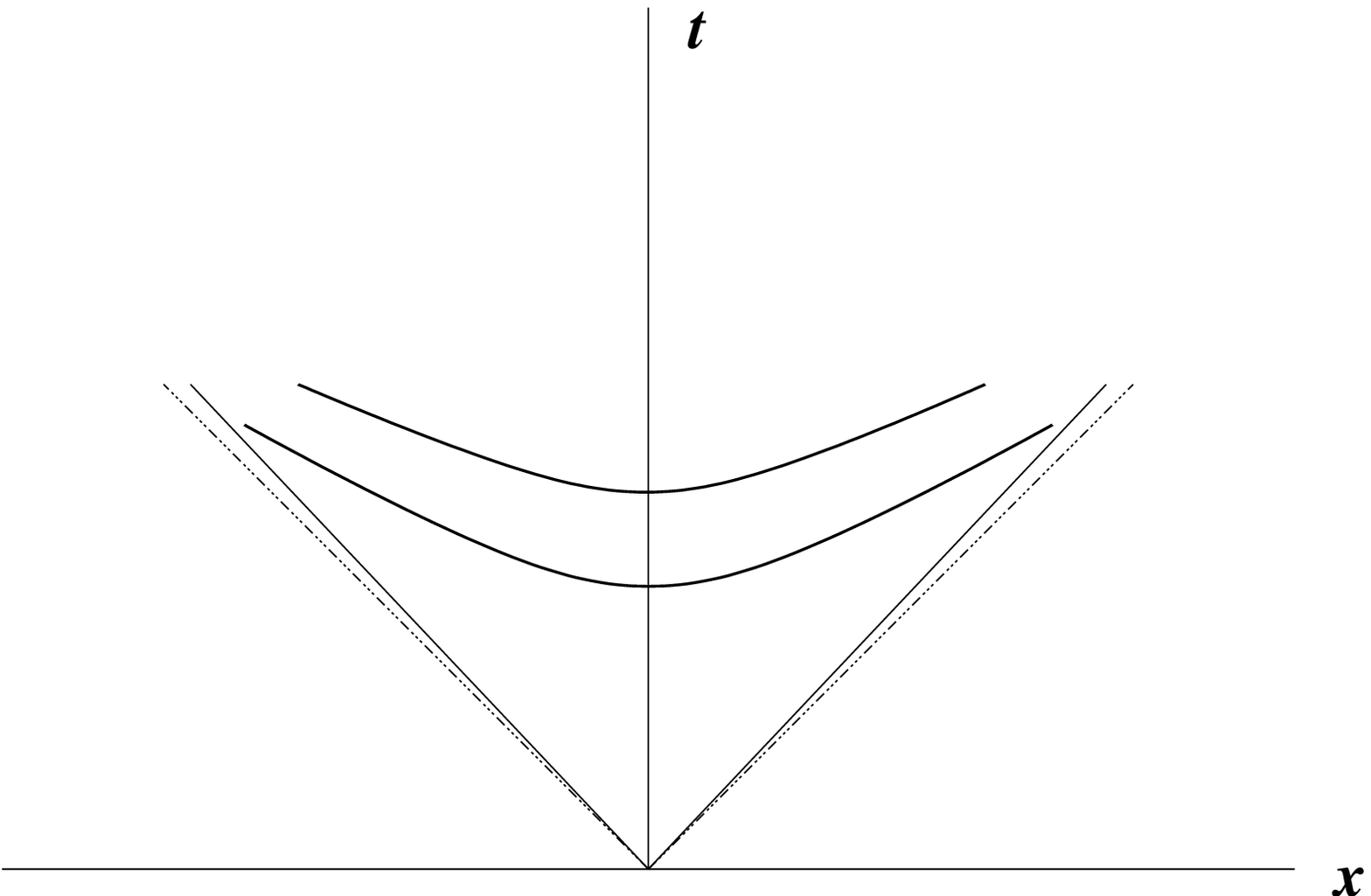} & 
\includegraphics[width=45mm]{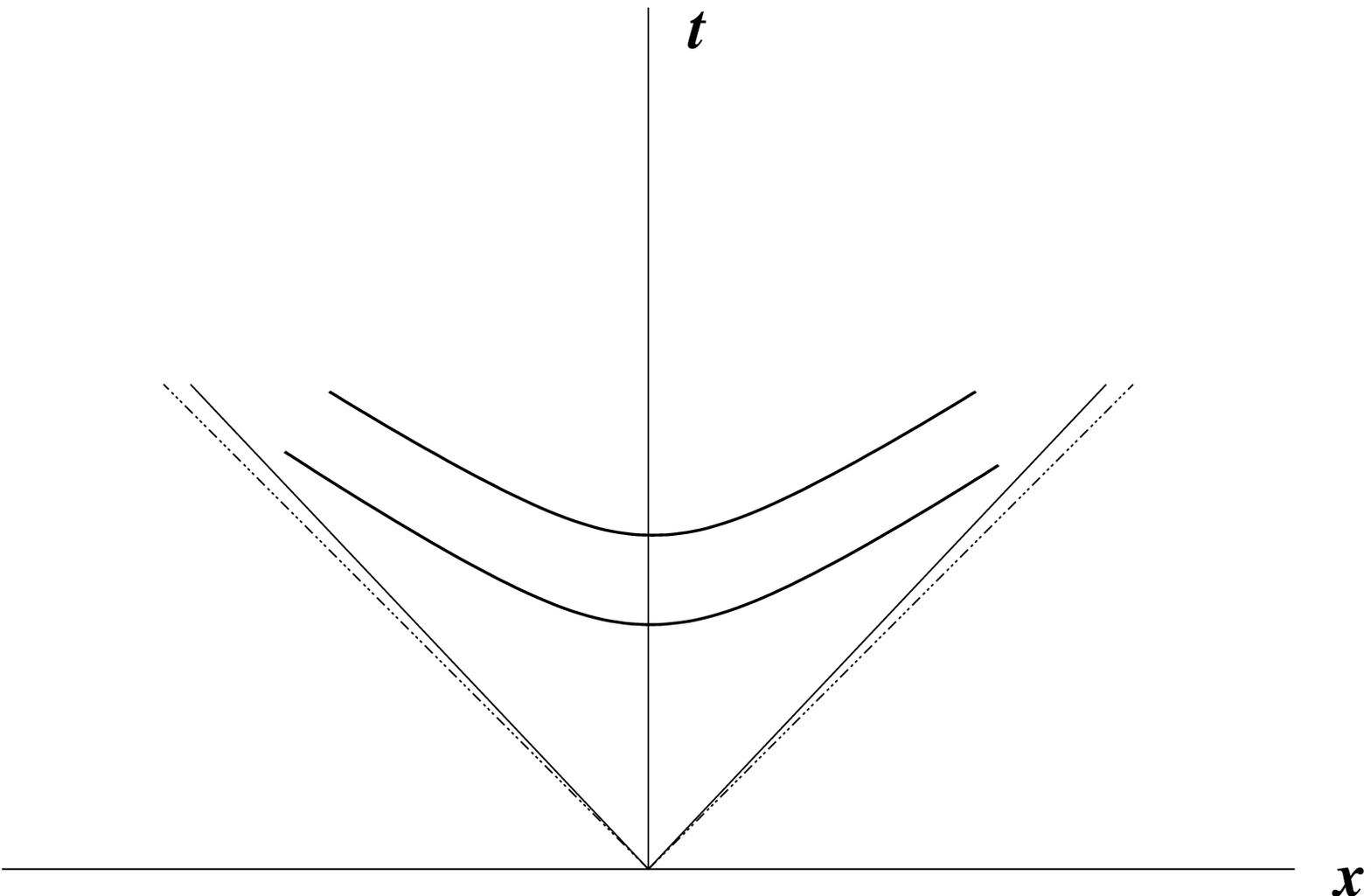} \\
Entropy & Mass \\
[5mm]
\includegraphics[width=45mm]{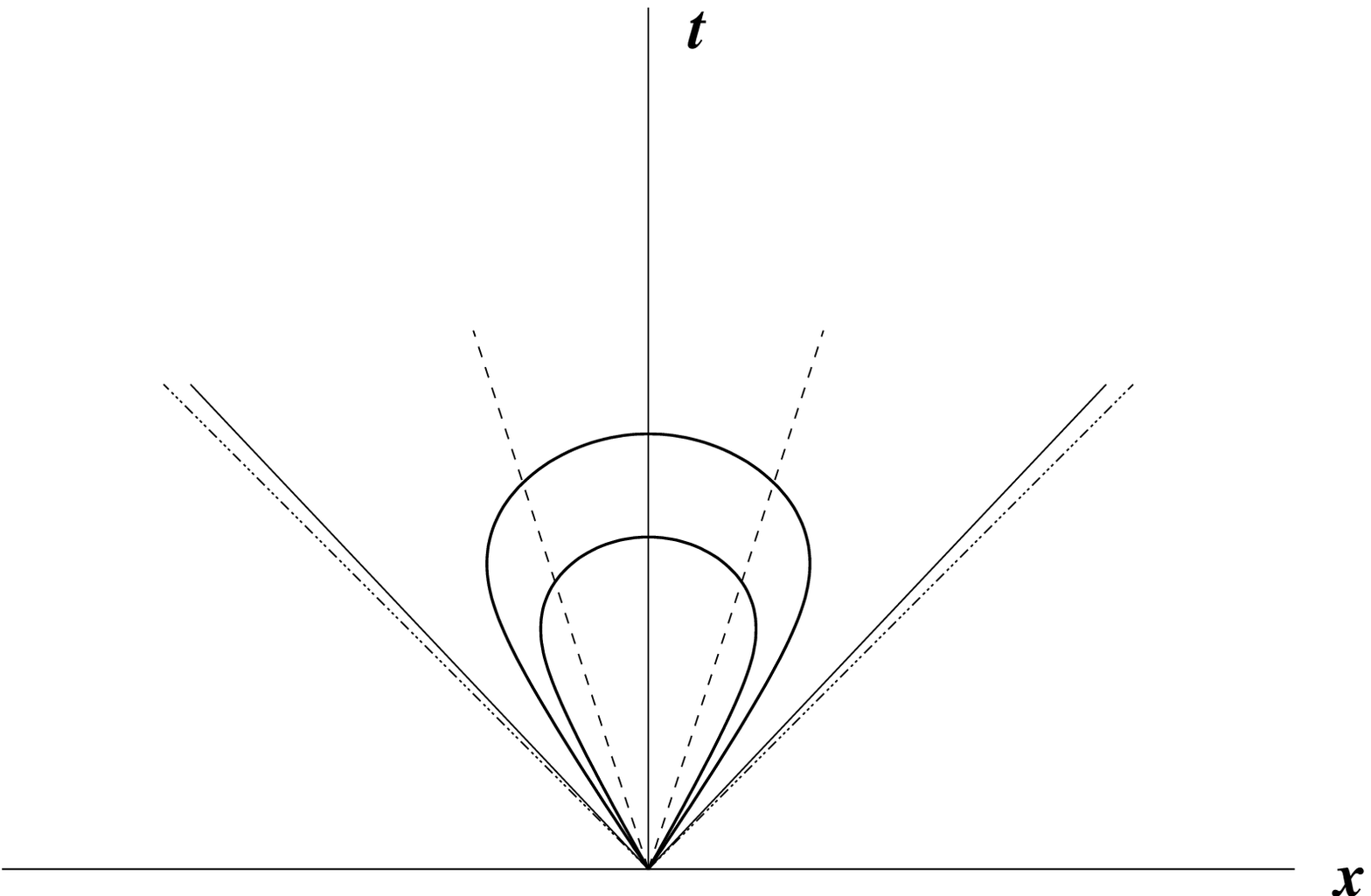} & 
\includegraphics[width=45mm]{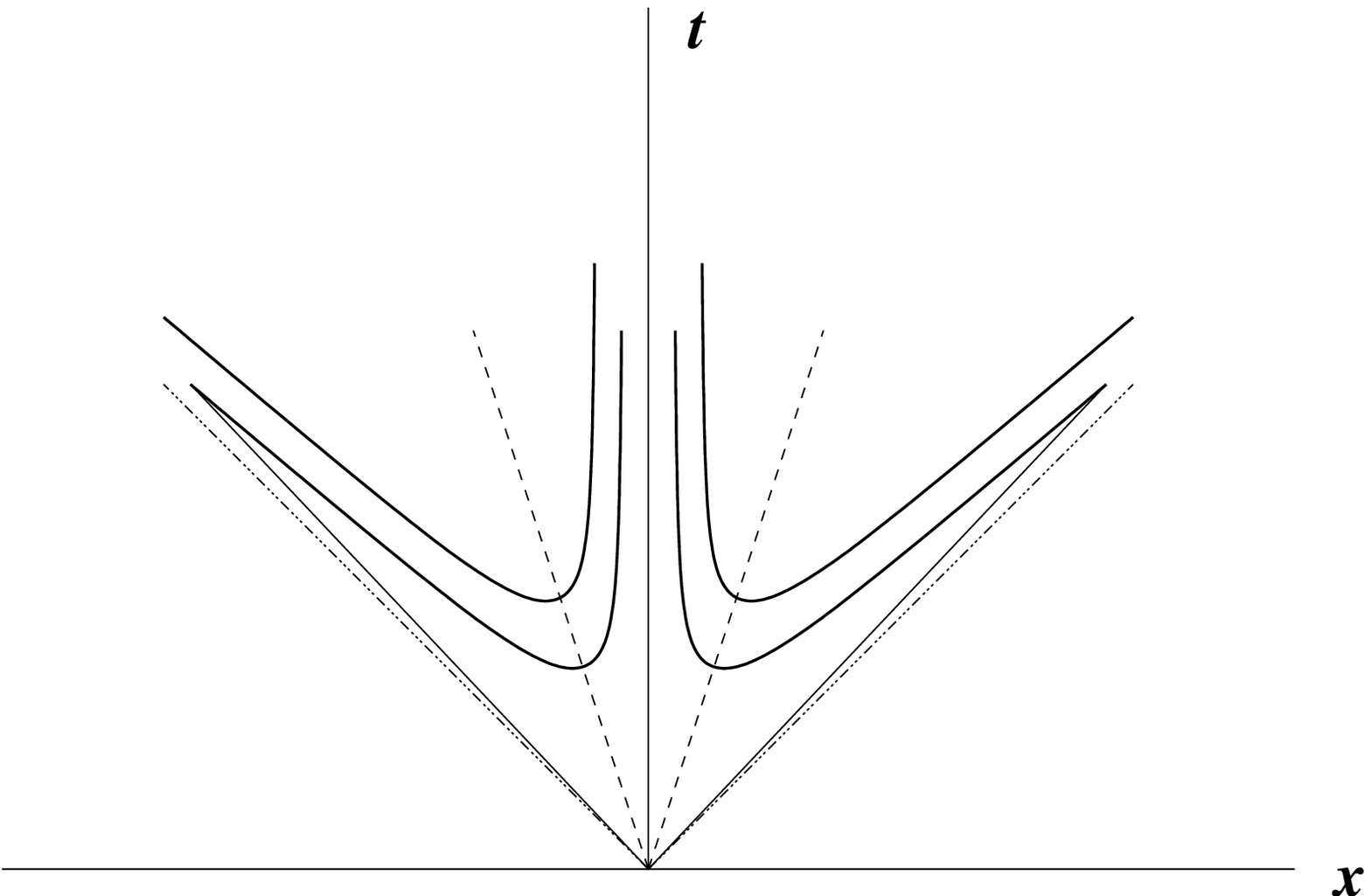}  \\
Temperature & Charge

\end{tabular}
\end{center}

\caption{The state space of the Reissner-Nordstr\"om black hole is
a wedge inside the forwards light cone of a 1+1 dimensional
Min\-kow\-ski space. We show curves of constant entropy (spacelike
hyperbolas), constant mass (also spacelike), constant temperature, and
constant charge. The latter two become null at the ``Davies point'',
$Q/M=\sqrt{3}/2$ \cite{Davies} which is given by a
dashed line of constant electric potential.}

\end{figure}

\section{Kerr black hole, spacetime dimension ${\bf\it d}$}

In higher dimensional spacetimes there can be more than one angular
momentum. For the Kerr black hole with a single nonzero spin we
obtain the event horizon from \cite{Myers}

\be{\label{eq:KerrHorizon}
r_+^{\,\,2} + a^2 - \frac{\mu}{r_+^{\,\,d-5}} = 0.
}

The area of the event horizon is given by
$A = \Omega_{(d-2)} r_+^{\,\,d-4} \: (r_+^{\,\,2} + a^2 )$
where $\Omega_d$ is the area of the unit $d$-sphere.
The ADM mass of the hole is defined by $\mu = \frac{4 M}{d-2}$. The
angular momentum per unit mass is
$a = \frac{d-2}{2}\frac{J}{M}$.
We have $S=\frac{kA}{4G}$ and set $\dis k = 1/\pi$ and set Newton's
constant $G$ to $\Omega_{(d-2)}/4\pi$ to obtain the entropy
$S = r_+^{\,\,d-4} (r_+^{\,\,2} + a^2) = r_{+} \mu.$
Explicit $S$ in arbitrary dimension cannot be obtained,
but we can work via the Weinhold metric. The mass is

\be{
\label{eq:Kerr-Mass}
M = \frac{d-2}{4}S^{\frac{d-3}{d-2}} \paren{1 + \frac{4J^2}{S^2}}^{1/(d-2)}.
}

The Weinhold metric $g^W_{ij} = \partial_i \partial_j M(S, J)$ of the
$d$-dimensional Kerr black hole is found to be flat! In a diagonal
form \cite{ourpaper2}:

\be{
ds^2_W = - d\tau^2
+ {\frac{2(d-3)}{(d-2)}}
{\frac{(1- 4{\frac{d-5}{d-3}}u^2)}{(1 +4u^2)^2}}\tau^2 du^2.
}

\noindent $u = \onehalf \sinh 2\sigma$ gives Rindler coordinates for $d=4$:
$ds^2_W = -d\tau^2 + \tau^2 d\sigma^2$. For $d=4$ the extremal limit is
$J/M^2 = 1$, hence $u$ is bounded by
$\abs{u} \leq \onehalf  \Leftrightarrow \abs{\sigma}
\leq \frac{1}{2} \sinh^{-1}1 \approx 0.4406$.
In Min\-kow\-ski style coordinates the state
space of the Kerr black hole is a wedge bounded by
$\abs{\frac{x}{t}} \leq \sqrt{\frac{\sqrt{2}-1}{\sqrt{2} + 1}}$.
For $d=5$ the extremal limit is given by $J^2/M^3 = 16/27$ and
$-\infty \leq u \leq  \infty $ where
$u = \onehalf \tan \sqrt{3}\sigma$. Hence we obtain a wedge with
wider opening angle $\abs{\sigma} \leq \frac{1}{\sqrt{3}}\arctan \infty
= \frac{\pi}{2\sqrt{3}} \approx 0.9069$.

Remarkably, the wedge of the $d \geq 6$ Kerr black
hole fills the entire light cone. This is because for black holes in
$d \geq 6$ there are no extremal limits.

\begin{figure}[t]
\begin{center}

\includegraphics[width=60mm]{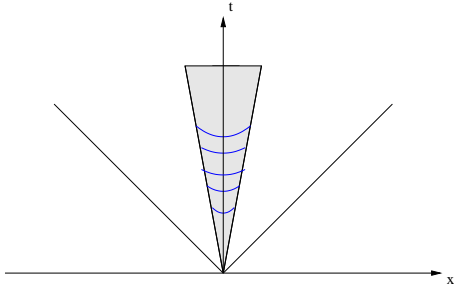}

\end{center}

\caption{The state space of $d=4$ Kerr black holes shown as
a wedge in a flat Min\-kow\-ski space. The slope of the wedge measures
approximately 80$^\circ$ from the $x$-axis. Curves of constant entropy
give causal structure to the state space of the black hole.}

\end{figure}

In any dimension, we obtain the Ruppeiner metric from the Weinhold
metric using the conformal relation. It is a curved metric with the
curvature scalar

\be{
\dis
\label{eq:Kerr-curvature}
\mathcal{R} = - \frac{1}{S}
\frac{\dis 1 - 12\frac{d-5}{d-3}\frac{J^2}{S^2}}
{\dis \paren{1 - 4\frac{d-5}{d-3} \frac{J^2}{S^2}}
\paren{1 + 4\frac{d-5}{d-3} \frac{J^2}{S^2}}}.
}

In $d=4$ the curvature scalar diverges along the curve $4J^2 = S^2$.
In $d=5$ the curvature is reduced to $\mathcal{R} = -\frac{1}{S}$
which diverges in the extremal limit of the $d=5$ Kerr black hole.

For $d \geq 6$ we have a curvature blow-up but not in the limit of
extremal black hole, rather at $4J^2 = \frac{d-3}{d-5} S^2$.
This is where Emparan and Myers \cite{Emparan-Myers} suggest
that the Kerr black hole becomes unstable and changes its behavior
to be like a black membrane.

Note that in $d=5$ for some values of the parameters, there exist
``black ring'' solutions \cite{Emparan-Reall}
whose entropy is larger than that of the black hole studied in our
papers. A careful observation of curvature scalar $\mathcal{R} =
-\frac{1}{S}$ indicates that nothing special happens to the Gibbs
surface of the Kerr black hole \emph{a la} Myers-Perry.

\section{Our flatness theorem}

\noindent We can now state a small theorem, namely: If $\psi(x,y) =
x^af(y/x)$ then the information metric is flat
\cite{ourpaper3}. The converse does not hold.
The most straightforward way to prove the theorem is to change to the
new coordinates $\psi = x^af(x^by)$, and $\sigma = x^by$. An explicit
calculation shows that

\begin{eqnarray} ds^2 &=& \left( \frac{a-1}{a} - \frac{b(b+1)}{a^2} 
\frac{\sigma f^\prime}{f}\right) \frac{d\psi^2}{\psi}
+ 2(b+1)\left( \frac{1}{a}\frac{f^\prime}{f} \frac{b}{a^2}
+ \frac{\sigma f^{\prime 2}}{f^2}\right) d\psi d\sigma \nonumber \\
&+& \psi 
\left( \frac{f^{\prime \prime}}{f} - \frac{2b + a + 1}{a}\frac{f^{\prime 2}}
{f^2} -  \frac{b(b+1)}{a^2}\frac{\sigma f^{\prime 3}}
{f^3}\right) d\sigma^2 \ .
\end{eqnarray}

\noindent This is diagonal if $b = -1$. If we introduce the new
coordinate $r = \sqrt{\psi}$ it is also manifestly a flat metric, and
it covers a wedge shaped region. Given the function $f$ we can
reparametrize $\sigma$ so that we end up with polar coordinates, or
Rindler coordinates if the metric is Lorentzian, and read off the
opening angle of the wedge. Anyway the small theorem is proved. There
is an exception if $b = - 1$ and $a = 1$, since then the metric is
degenerate.

We can restate our small theorem in thermodynamical language: Let $S =
M^af(M^bQ)$. If $b = - 1$ the Ruppeiner metric is flat. If $a+b = 0$
the Weinhold metric is flat.

An explicit example: The Reissner--Nordstr\"om black hole in arbitrary
spacetime dimension $d$ has the fundamental relation\ 
$S = M^c\left( 1 + \sqrt{1 - \frac{c}{2}\frac{Q^2}{M^2}}\right)^c$ \ ,
$c \equiv \frac{d-2}{d-3}$.
The Ruppeiner geometry is a timelike wedge in a flat
Min\-kow\-ski space, with an opening angle that grows with $d$. It is
a black hole if $d \geq 4$.

This theorem is now helping us treating other sorts of black holes as
e.g.\ dilaton black holes, see \cite{ourpaper4,narit2007} and
forthcoming papers.

For computer algebra calculations we used {\sc GRTENSOR} for {\sc
MAPLE} and {\sc CLASSI}.

Several interesting papers has recently been published, see for
instance Arcioni \& Lozano-Tellechea \cite{arcioni}, Shen \etal
\cite{shen}, Quevedo \cite{quevedo}, Mirza \& Zamani-Nasab
\cite{mirza} and Ruppeiner \cite{ruppeiner3}.

\section{Summary of results}

\begin{center}
\begin{table*}[ht]
\begin{tabular}{|c|l|l|l|}
\hline\hline
\textbf{Spacetime dimension} & \textbf{Black hole family} & \textbf{Ruppeiner}
 & \textbf{Weinhold}  \\
\hline
any $d \geq 4 $ & Kerr                  &  Curved & Flat    \\ \cline{2-4}
                & Reissner--Nordstr\"om &  Flat   & Curved  \\ \hline 
$d=4$           & Kerr-Newman           &  Curved & Curved  \\ \cline{2-4}
                & RN-anti-de Sitter     &  Curved & Curved  \\ \hline 
$d=5$           & double-spin Kerr      &  Curved & Curved  \\ \cline{2-4} 
\hline\hline
\end{tabular}
\caption{Geometry of black hole thermodynamics.}
\end{table*}
\end{center}

\vspace{-8mm}

\section*{Acknowledgments}

We thank Ingemar Bengtsson for valuable discussions.

\end{document}